\documentclass{article}

\usepackage[utf8]{inputenc} 
\usepackage{microtype}
\usepackage{graphicx}
\usepackage{subfigure}
\usepackage{booktabs} 

\usepackage{hyperref}

\usepackage[accepted]{icml2020}

\icmltitlerunning{Automatic segmentation for prediction of tuberculosis}

\begin{document}

\twocolumn[
\icmltitle{Automatic semantic segmentation for prediction of \\ tuberculosis using lens-free microscopy images}



\icmlsetsymbol{equal}{*}

\begin{icmlauthorlist}
\icmlauthor{Dennis Núñez-Fernández}{one}
\icmlauthor{Lamberto Ballan}{two}
\icmlauthor{Gabriel Jiménez-Avalos}{one}
\icmlauthor{Jorge Coronel}{one}
\icmlauthor{Mirko Zimic}{one}
\end{icmlauthorlist}

\icmlaffiliation{one}{Laboratorio de Bioinformática y Biología Molecular, Universidad Peruana Cayetano Heredia, Peru}
\icmlaffiliation{two}{Visual Intelligence and Machine Perception Group,
University of Padova, Italy}

\icmlcorrespondingauthor{Mirko Zimic}{mirko.zimic@upch.pe}

\icmlkeywords{Machine Learning, ICML}

\vskip 0.3in
]



\printAffiliationsAndNotice{}  

\begin{abstract}
Tuberculosis (TB), caused by a germ called Mycobacterium tuberculosis, is one of the most serious public health problems in Peru and the world. The development of this project seeks to facilitate and automate the diagnosis of tuberculosis by the MODS method and using lens-free microscopy, due they are easier to calibrate and easier to use (by untrained personnel) in comparison with lens microscopy. Thus, we employ a U-Net network in our collected dataset to perform the automatic segmentation of the TB cords in order to predict tuberculosis. Our initial results show promising evidence for automatic segmentation of TB cords.
\end{abstract}

\section{Introduction}

Around 1.7 billion people, equivalent to a quarter of the world’s population, are infected with Mycobacterium tuberculosis; 5–10\% of them will develop TB disease during their lifetime. Also, it is currently estimated that 1.2 million people died from tuberculosis in 2018. \cite{1_}.

The MODS method (Microscopic Observation Drug Susceptibility Assay) \cite{2_} developed in Peru allows the growth and recognition of morphological patterns of mycobacteria in a liquid medium and quickly, directly from a sputum sample, is a cost-effective option, since in just 7 to 21 days TB can be detected, which is a fast and low-cost alternative test. This method has recently been added to the list of rapid tests authorized by the National Tuberculosis Prevention and Control Strategy. In addition, complementing the MODS method with lens-less microscopy is very positive because it is easier to use and to calibrate by no trained personnel.

In recent years Convolutional Neural Networks (CNNs), a class of deep neural networks, have become the state-of-the-art for object recognition in computer vision \cite{3_}, and have high potential in object detection \cite{4_,5_} and image segmentation \cite{3_}.

For the analysis of medical images via deep learning, the segmentation of the object has been often essential. Manual segmentation by experienced clinicians, is important; however, it is laborious and time-consuming, and may be subjective. Another approach for image segmentation, a convolutional neural network known as U-net, has recently shown promising results \cite{6_}. U-net has been applied for some medical and microscopy images, including brain tissue characterization and segmentation, vessel wall segmentation, detection and counting of cells, and identification of bacterias \citep{10_,8_,7_,11_}.

The purpose of this work is to evaluate the feasibility of fully automatic TB cords segmentation on lens-free images. We applied U-net for the automatic segmentation task.

\section{Methodology}

The proposed methodology is based on making use of an automatic segmenting of Tuberculosis (TB) cords using a convolutional neural network in order to have a more robust and efficient automatic segmenting device compared to traditional segmentation techniques. Therefore the proposed methodology has the following steps: First the total image of 3840x2700 pixels is divided into sub-images of 256x256 pixels, then each sub-image is processed through the proposed deep neural network, which gives the automatic segmentation of TB cords for in each sub-image. Later, these sub-images are concatenated to reconstruct the automatic segmentation of TB cords for the full image.

\begin{figure}[H]
\vskip 0.2in
\begin{center}
\centerline{\includegraphics[width=\columnwidth]{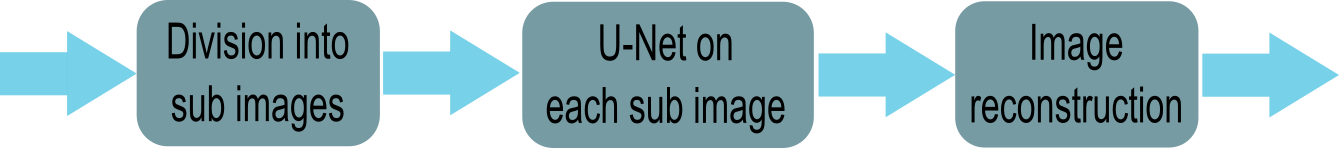}}
\caption{Overview of the methodology.}
\label{overview}
\end{center}
\vskip -0.2in
\end{figure}

The images were collected by the iPRASENSE Cytonote lens-free microscope. These images were reconstructed from the holograms that were obtained by the lens-free microscope. At data collection, we obtain 10 full images with a dimension of 3840x2700 pixels each one and presented in grayscale format.

\begin{figure}[H]
\vskip 0.2in
\begin{center}
\centerline{\includegraphics[width=0.40\columnwidth]{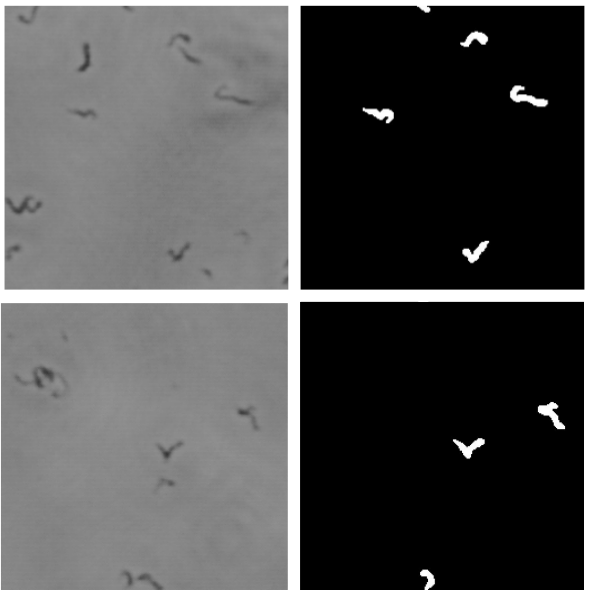}}
\caption{Samples of the generated dataset.}
\label{samples}
\end{center}
\vskip -0.2in
\end{figure}

For the generation of the dataset, the following steps were carried out: First a medical expert (with extensive experience in TB research) highlighted the TB cords in a rectangle. Then, a staff, which was trained to recognize the morphology of TB cords, delineated the edges of the TB cords. After, the sub-images and their TB cord masks were obtained using simple computer vision techniques. The generated dataset has 150 grayscale and 150 binary sub-images of 512x512 pixels size, see Figure~\ref{samples}.

\begin{figure}[H]
\vskip 0.2in
\begin{center}
\centerline{\includegraphics[width=0.85\columnwidth]{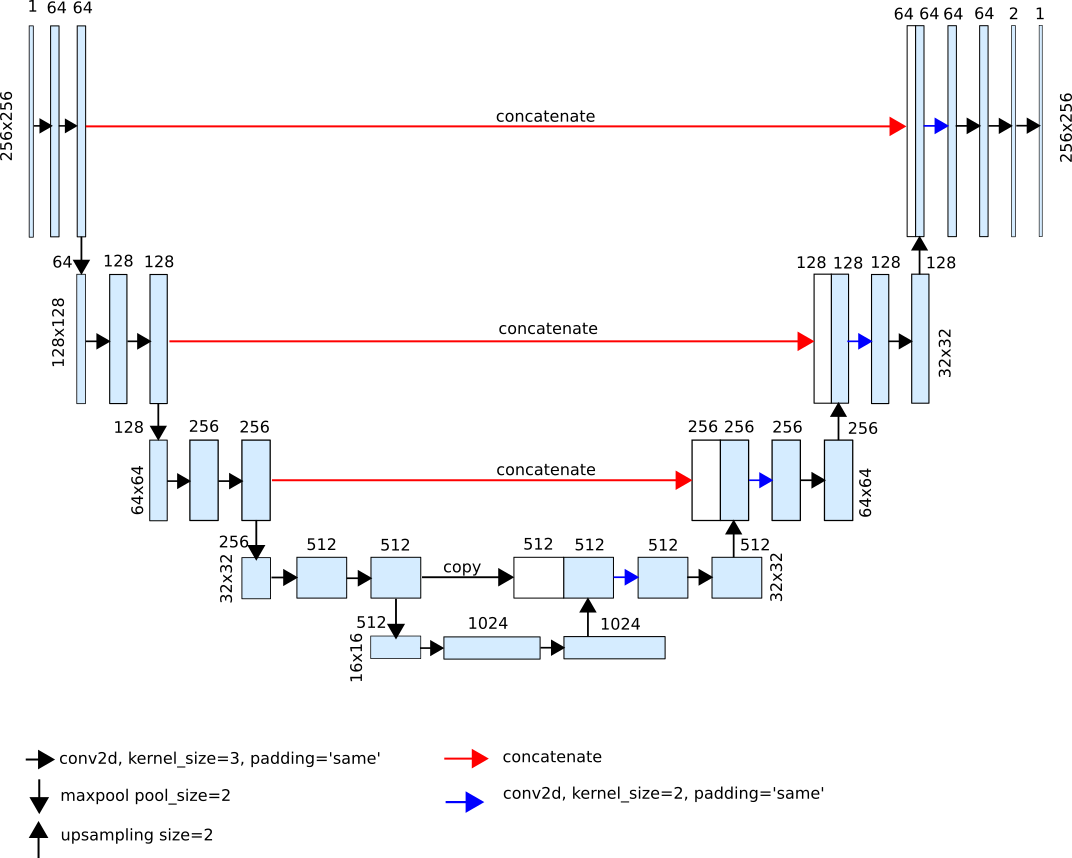}}
\caption{The U-Net architecture used for segmentation.}
\label{unet}
\end{center}
\vskip -0.2in
\end{figure}

We make use of the U-Net architecture and some hyper parameters were modified in order to work with our collected dataset and to produce an accurate segmentation. As Figure~\ref{unet} shows, the U-Net consists of a contract and an expansive path. During the contraction, the spatial information is reduced while feature information is increased. The expansive path combines the feature and spatial information.

\section{Results}

\begin{figure}[H]
\vskip 0.2in
\begin{center}
\centerline{\includegraphics[width=0.80\columnwidth]{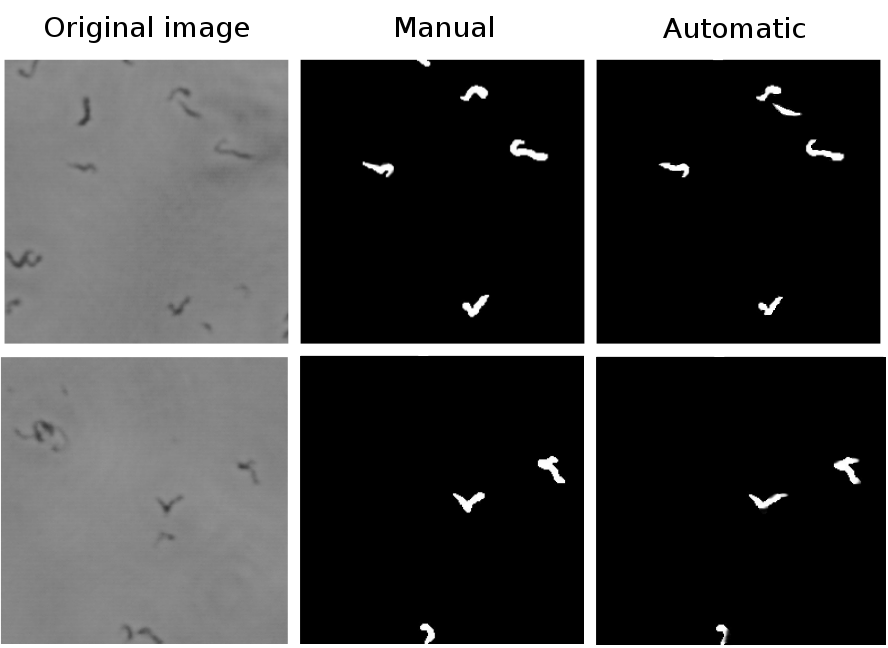}}
\caption{Results of the automatic segmentation.}
\label{results}
\end{center}
\vskip -0.2in
\end{figure}

For the evaluation of the U-net architecture, we used the next segmentation metrics: intersection-over-union (IoU or Jaccard Index) and pixel accuracy. We employed 120 sub-images for training the U-Net and 30 sub-images for testing. In this sense, we obtain an IoU of 0.88 and a pixel accuracy of 92.01\%. The results show a good performance despite the fact that the dataset is quite noisy and has a resolution that makes difficult to define the TB cords. As you can see in Figure~\ref{results}, the automatic segmentation looks quite similar to the manual segmentation. Some regions of the automatic segmentation are wrong segmented due the low resolution and the difficulty to define Tb cords in such areas.

\section{Conclusions}

In this work we demonstrated that our system is capable for automatic segmentation of TB cords with an IoU of 0.88 and a pixel accuracy of 92.01\%. This by using images captured from a lens-free microscopy and using the U-net architecture. These initial tests show promising results despite the fact that the project is still under development, i.e. more samples are still being collected and some improvements are still being implemented. The most important fact is that this project contributes to a fast, cost-effective, and globally useful tool for tuberculosis detection, especially in the Peruvian rural areas where medical resources are limited.

\section*{Acknowledgements}

This project has been financed by CONCYTEC, by its executing entity FONDECYT, with the objective of promoting the exchange of knowledge between foreign academic institutions and Peru.

 \clearpage

\bibliography{references}
\bibliographystyle{icml2020}

\end{document}